\documentclass[journal]{IEEEtran}

\usepackage{amssymb}
\usepackage{amsmath}
\usepackage{makecell}
\usepackage{cite}
\usepackage{graphicx}
\usepackage{url}
\usepackage{array}
\usepackage{algorithmic}
\usepackage{algorithm}
\usepackage{multirow}
\usepackage{graphics}
\usepackage{color}
\usepackage{enumitem}
\begin{document}

\title{ Classification of COVID-19 from CXR Images in a 15-class Scenario: an Attempt to Avoid Bias in the System}
\author{Chinmoy Bose and Anirvan Basu  
	
	\IEEEcompsocitemizethanks{
		\IEEEcompsocthanksitem  
		\IEEEcompsocthanksitem Chinmoy Bose is with TwoPiRadian Infotech LLC, USA (email: chinmoy.bose@gmail.com)
		\IEEEcompsocthanksitem Anirvan Basu is with Accenture Applied Intelligence, France (email: anirvan.basu@insead.edu)		
	}}

\maketitle

\begin{abstract}
As of June 2021, the World Health Organization (WHO) has reported 171.7 million confirmed cases including 3,698,621 deaths from COVID-19. Detecting COVID-19 and other lung diseases from Chest X-Ray (CXR) images can be very effective for emergency diagnosis and treatment as CXR is fast and cheap. The objective of this study is to develop a system capable of detecting COVID-19 along with 14 other lung diseases from CXRs in a fair and unbiased manner.   The proposed system consists of a CXR image selection technique and a  deep learning based model to classify 15 diseases including COVID-19. The proposed CXR selection technique aims to retain the maximum variation uniformly and eliminate poor quality CXRs with the goal of reducing the training dataset size without compromising classifier accuracy. More importantly, it reduces the often hidden bias and unfairness in decision making. The proposed solution exhibits a promising COVID-19 detection scheme in a more realistic situation than most existing studies as it deals with 15 lung diseases together. We hope the proposed method will have wider adoption in medical image classification and other related fields.
\end{abstract}


\section{Introduction}\label{SecIntroduction}
Coronavirus Disease 2019 (COVID-19) is a new coronavirus that has not been previously identified. It can infect human beings causing illness varying from the very mild cold to severe health conditions including death. The immediate risk of being infected by COVID-19 is increasing day by day as the outbreak during the first wave affected almost the whole world. The second wave in 2021 infected many more people.  The World Health Organization (WHO) has reported 171.7 million confirmed cases including 3,698,621 deaths from COVID-19 as of June 4th, 2021. There is a tremendous demand for fast and easy to build accurate diagnostic systems during the pandemic  to compensate for shortage of medical staff and facilities, particularly in smaller hospitals and countries with limited healthcare infrastructure, particularly for rural areas. 

Chest X-Ray (CXR) is one of the diagnostic tools used by doctors. Typically, CXRs are analysed by radiologists. Due to the increasing number of COVID-19 cases, it is getting harder for radiologists to keep up with this demand. Proper understanding of COVID-19 in CXR is a challenging task even for experienced radiologists as the ability to confidently diagnose and accurately document the findings quickly can be unreliable \cite{GhoshalTucker2020}. This justifies the significant demand for the development of computer aided diagnosis systems. Detecting COVID-19 together with other lung diseases based on CXR processing can be challenging, and it is still under active exploration. Lung disease and COVID-19 detection using CXRs, have been analyzed in a number of studies \cite{FShi2020, DDong2020, LaureWynants2020}. Most of these studies are restricted to a small number of classes, and sometimes even to two classes: COVID and non-COVID.

Here we propose a methodology to combine a 14-class Chest X-Ray data set,  ChestX-ray14 dataset\cite{ChestXray14Dataset} together with COVID-19 X-Ray dataset to predict the probability of occurrence of the 15 different lung diseases including COVID-19 using a relatively small number of CXRs. Consideration of 15 classes together makes the problem very challenging.  A deep learning based CXR classifier is proposed to predict the probability of 15 different classes including COVID-19. We propose a method to select the CXRs for designing the system which attempts to minimize the bias in decision making. The performance of the proposed system is satisfactory. However, to the best of the authors' knowledge no study has been conducted to optimally select CXR images for learning the classifiers.

\section{Related Work}\label{SecRelatedWork}

The chest X-ray (CXR) remains the most commonly ordered imaging study in medicine and is no different for COVID-19. Several studies have been published recently on COVID-19 detection using CXR utilizing Deep Learning based techniques \cite{FShi2020, DDong2020, LaureWynants2020}.  It is important to note that the COVID-19 dataset continues to evolve as new patient cases are continuously added and are made available publicly on a regular basis. This motivates us to focus on selection of CXR for detecting COVID-19 along with other lung diseases. 

Rajpurkar et al. \cite{PranavRajpurkar2017} proposed a DenseNet-121 based Deep Learning (DL) model to classify 14 different thoracic diseases, including pneumonia  based on the ChestX-ray14 dataset. They reported that the model exceeds the average radiologist's performance on the pneumonia detection task \cite{PranavRajpurkar2017}.
Inspired by the results reported in Rajpurkar et al. \cite{PranavRajpurkar2017} , Cohen et al. \cite{JosephPaulCohen2020} also used the DenseNet-121 architecture \cite{DenseNet}.  The models were trained using the Chest-Xray14 dataset\cite{ChestXray14Dataset}. The performance was compared with that of Weng et al. \cite{XinyuWeng2017}, which exhibits comparable performance with that of the original paper. A web application Chester \cite{JosephPaulCohen2020} was introduced for disease prediction using CXR.  Haghanifar et al. \cite{ArmanHaghanifar2020}  proposed a network named COVID-CXNet also based on DenseNet-121, but they  focused on distinguishing between COVID-19 and normal lung CXR images as well as a three-class problem: COVID pneumonia, non-COVID pneumonia, and normal.  

Computed Tomography (CT) images of the lungs have also been used for detection of COVID-19.  In \cite{XXu2020} authors explored lung CT images to differentiate COVID-19 from Influenza-A Viral Pneumonia and healthy cases using 3D CNN based deep learning \cite{XXu2020}. They have reported an overall accuracy of 86.7\%. They suggested that their system could be a promising supplementary diagnostic method for frontline clinical doctors. This study lacks differentiation of COVID-19 from other diseases such as pneumonia that a patient might have. Developing a system which can distinguish between COVID-19 CXRs and other similar looking CXRs will definitely help frontline medical staff.
Narin et al \cite{ANarin2020} experimented with CXR to differentiate COVID-19 from normal cases using different Deep Learning (DL) models: InceptionV3, ResNet50, and  InceptionResNetV2. They concluded that ResNet50 is the best choice for the dataset they considered. They  reported an accuracy of 98\%. 

In \cite{IDApostolopoulos2020} Apostolopoulos et al.  experimented on limited COVID-19 data to choose the base DL model for transfer learning. They concluded that VGG19 \cite{VGG} performed the best followed by Mobile Net \cite{MobileNet}. The other deep networks considered in their study included Inception \cite{InceptionNetV3}, Xception \cite{Xception} and  InceptionResNetV2 \cite{InceptionNetV4}. They considered both the  3-class  (Covid-19, Pneumonia, and Normal) and the two-class  (Covid-19 and Non Covid-19) classification problems. In both cases VGG19 performed the best \cite{IDApostolopoulos2020}. A VGG16 based CNN   has also been proposed  in \cite{LawrenceOHall2020}. Authors used the 10-fold cross validation  on the dataset from Cohen \cite{JPCohen2020}. A data augmentation strategy was employed to increase the size of the dataset. The proposed approach achieved an overall accuracy of 96.1\%.
Authors in \cite{JZhang2020} investigated  differentiation of viral pneumonia from non-viral pneumonia. Viral pneumonia can be caused by Influenza A/B viruses, respiratory syncytial virus, coronaviruses, herpes simplex, measles, chickenpox, and more seriously, by some novel viruses.  They proposed a confidence-aware anomaly detection (CAAD) model to distinguish viral pneumonia cases from non-viral pneumonia cases and healthy controls using chest X-rays and reported 96\% sensitivity on COVID-19 cases. 

Zhang et al. \cite{JianpengZhang2020} also proposed a CAAD system, that works by chaining a convolutional feature detector, an anomaly detection module and a confidence prediction module. They formulate discrimination of viral-pneumonia from non-viral pneumonia and healthy control as an anomaly detection (i.e., one-class classification) problem.  CAAD utilizes an anomaly detection module that enables it to potentially train in the absence of viral pneumonia CXR cases. 

A CNN based COVID-Net architecture was built using generative synthesis in \cite{LWang2020}. Here five CXR repositories were combined to classify COVID-19 pneumonia, non COVID-19 pneumonia and no pneumonia. The reported highest accuracy of the system was 93.3\% surpassing results obtained from VGG19 and  ResNet-50.
Another scheme to distinguish CXR images of COVID-19 patients from those of normal or those infected by other viral or bacterial diseases has been proposed using texture features in \cite{DAlKarawi2020}. Authors reported  that LBP-Gabor filters attained the highest and most stable accuracy.  Use of hierarchical clustering to classify pneumonia into various clusters including  COVID-19, SARS, and MERS, has also been tried \cite{RMPereira2020}.  

 Rajaraman et al. \cite{SRajaraman2020} used several pretrained CNNs and further trained those using publicly available CXRs and the best performing networks are pruned to reduce the complexity of the networks. The ensemble of the networks is used for classification of  CXR images as COVID-19 Pneumonia, Bacterial Pneumonia, Pneumonia of Unknown Type, and Normal. A CNN named COVID-NET has been proposed in \cite{LWang2020_DUP}  to distinguish between COVID-19 and Non-COVID-19 cases using CXR images.  
  A semi-supervised deep learning system based on Auto-Encoders, named CoroNet, for the diagnosis of COVID-19 from CXRs has been developed in \cite{SKhobahi2020}. The authors used three open access datasets for experiments and reported an overall accuracy of 93.5\%.

Ferhat et al. \cite{FerhatUcar2020} proposed a novel method based on deep Bayes-SqueezeNet, called COVIDiagnosis-Net, to classify Normal (no infection), Pneumonia (bacterial or non-COVID viral infection) and Covid (COVID-19 viral infection). To overcome the data imbalance problem, an offline data augmentation involving noise, shear and  brightness was used. They reported 98.3\% accuracy. We note here the 15 class problem that we deal with in this investigation is a much more difficult one.

In \cite{AsmaaAbbas2020},  a deep CNN called DeTraC has been proposed. This method can deal with irregularities in the image dataset by analysing its class boundaries using a class decomposition mechanism. They considered normal and COVID-19 CXRs for a two-class problem. To generate more instances a data augmentation technique was applied and histogram modification was applied to enhance the images.
Like several other studies, SUK et al. \cite{BukhariSUK2020} considered the three-class problem:  normal, pneumonia and COVID-19.  Here, a ResNet-50-based model has been used yielding an  accuracy of 98.18\%. A mobile app named MobileXpert based on a deep CNN has also been developed to distinguish between COVID-19, viral pneumonia and normal cases using CXR images\cite{XLi2020}. For a robust performance, authors use novel loss functions for the training.
Hassanien et al. \cite{AEHassanien2020} proposed a classification system that uses multi-level thresholding and an SVM to detect COVID-19 in CXR images. 
Arpan et al. \cite{ArpanMangal2020} proposed CovidAID, a CheXNet based model for transfer learning. CheXNet is based on DenseNet-121 which can classify 14 classes. They replaced CheXNet’s  last layer of 14 classes with 4 Sigmoid units to classify Normal Lung, Bacterial Pneumonia, Viral Pneumonia and COVID-19. They reported  a precision of 87.2\% in this 4-class problem. 

Pedro et al. \cite{PedroRASBassi2020} have also used a DenseNet-121 based system to classify Normal, Viral Pneumonia and Covid-19 from CXRs. They fine-tuned the ImageNet based pre-trained model with the NIH ChestX-Ray14 dataset and then with Covid-19 dataset. They reported a test accuracy of 99.4\%.
Castiglioni et al. \cite{ICastiglioni2020}, on the other hand,  trained an ensemble of 10 ResNets and achieved a ROC-AUC of 0.80 for binary classification Covid-19. 
A DesnseNet-121 based architecture has also been used in \cite{IslamMT2020} to classify CXR images as L and H phenotypes. H-type patients have pneumonia-like thickening of the lungs and require ventilation to survive. Their objective was to help the treatment plan as the number of ventilators are limited. The fine-tuned network showed a sensitivity of 80\% for COVID-19 detection and 93\% for pneumonia detection.
A ResNet-50-based system is proposed in \cite{KanaGEB2020}  to classify CXR as healthy individuals, bacterial and viral pneumonia, and COVID-19 positives. 
Rezaul  et al. \cite{MdRezaulKarim2020} developed a DeepCOVIDExplainer for automatic detection of normal, pneumonia, and COVID-19 cases from CXR images. They used an ensemble based on DenseNet, ResNet, and VGGNet utilizing transfer learning. Models were trained on COVIDx dataset and Grad-CAM++ \cite{MdRezaulKarim2020} was used for explainability. 
 
CoroNet, a CNN  based on Xception \cite{SKhobahi2020} that can classify CXR as normal, bacterial, viral,  and COVID-19 was proposed in \cite{AsifIqbalKhan2020}.  The authors developed a balanced dataset to train and test their model. COVID-19 images were obtained from the open source GitHub repository \cite{JPCohen2020}.
Authors in \cite{RahulKumar2020} proposed a ResNet152 based model to classify CXRs as COVID-19, Pneumonia, and Normal. SMOTE \cite{Nitesh2002} was used for balancing the imbalanced data points of COVID-19 and Normal patients. 

In \cite{PierreGBMoutounetCartan2020} different networks were explored: VGG16, VGG19, InceptionResNetV2, InceptionV3 and Xception  followed by a flat multi-layer perceptron and a final 30\% drop-out. Authors concluded that VGG16 is the best performing network trained over 3 classes (COVID-19, No Finding, Other Pneumonia). The two sources of data  used were \cite{JPCohen2020} and  \cite{DSKermany2018}. They claimed that \cite{JPCohen2020} predominantly had data from ill patients while the other dataset \cite{KermanyDaniel2018} had healthy patients only.
A deep neural network based on Darknet-19, named  DarkCovidNet, was proposed in \cite{TOzturk2020}. The model was trained with data from \cite{JPCohen2020} and ChestX-ray8 database \cite{XiaosongWang2017}|. They reported that for the binary (covid-19 and no findings) and multi-class (covid-19, no findings, pneumonia) tasks the accuracy were 98.08\% and 87.02\%, respectively.

Rahimzadeh et al. \cite{MohammadRahimzadeh2020} proposed a CNN based on Xception and ResNet50V2. Features of Xception and ResNet50V2 are concatenated and then forwarded to a convolutional layer that is connected to a 3-class classifier. This system can classify CXR images as normal, pneumonia or Covid-19. Covid-19 CXRs were sourced from  Cohen et al.\cite{JPCohen2020}, other images were obtained from Kaggle RSNS pneumonia detection challenge. They reported an average accuracy of 99.56\%.

In this paper our objective is to classify Covid 19 CXR images along with 14 other lung diseases including pneumonia, pneumothorax, and fibrosis. This is a much more difficult problem than the covid-19 and non-Covid 19 classification, or most of the classification problems dealt with the methods discussed. However, compared to available CXR data for other thorax diseases, the number of CXR images available for Covid-19 patients is limited.  This would be the case even for any new future infection. Therefore, it would be very useful to augment the Covid-19 data set in a suitable manner so that successful algorithm for classification of Covid-19 along with other thorax diseases can be developed. This is what we have done here. We have proposed an innovative method to select images so that any bias in the designed system induced by the data set is minimized and this in turn makes the system a fair decision maker. As a feature extractor we have used DenseNet-121. The performance of the proposed system is satisfactory for the 15-class problem.

\section{Methodology}\label{SecMethodology}

\subsection{Datasets}\label{SecDataset}

\subsubsection{COVID-19 dataset \cite{JPCohen2020}}
This data set contains CXR images of COVID-19 patients that were collected from the publicly available GitHub repository built by Dr. Joseph Cohen \cite{JPCohen2020} not all of which are annotated, we have considered only the annotated ones.
This data set contains only 124 instances.

\subsubsection{ChestX-ray14 dataset [62]}
This data set includes 112,120 chest radiographs (frontal view) from   30,805  unique patients.   The data set covers patients suffering from 14 common thorax diseases. The 14 different thoracic pathologies considered are Atelectasis, Effusion, Infiltration, Cardiomegaly, Mass, Nodule, Pneumonia, Pneumothorax, Edema, Consolidation, Fibrosis, Emphysema, Pleural Thickening, and Hernia. This data set is labeled where the labels are generated from text-mining of the associated reports. Consequently, each image may have multiple labels. The labels generated from the text-mining of reports are estimated to be more than 90\% accurate.
Here, each image is of size  1024 x 1024 pixels. We could not find other studies that consider all these 15 classes together.

\subsection{Maximum Variation Image Selection}\label{SecMaximumVariationImageSelection}
\noindent The objective of this module is to select instances from different classes in such a manner so that there is balanced representation (to the extent possible) from each class and at the same time in the set of selected images the useful variation in the training set is maintained.
This will help to reduce chances of bias in the designed system. Moreover, images that may have derogatory impact during training would be dropped.
 

Figure 1, pictorially summarizes  the tasks involved in this module at a macro level.
\begin{figure}[!htb]
	\begin{center}
		\scalebox{0.3}[0.3]{\includegraphics{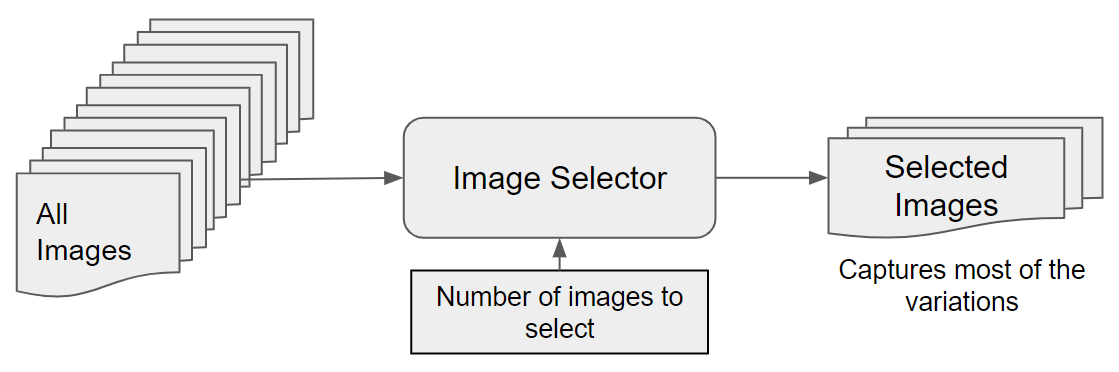}}
	\end{center}
	\caption{Objective of Maximum Variation Image Selection}
\end{figure}


For the Covid-19 data sets, since we have only 124 labeled images, we have considered them all. 
The number of instances from other classes are imbalanced. Use of such data sets, even though  a training-validation-test paradigm may exhibit good performance, may make the AI/ML system biased  to the most represented classes and thereby making "unfair" decisions for the other classes. The unfair behavior of an AI/ML system could be injected for other  reasons also. For a given disease, the variation in the CXRs may depend on several factors such as the devices used to obtain the CXRs, the prevalence of a disease and there could be many other factors influencing the characteristics of the CXRs. For example, in a sample there may be more tall patients compared to patients with short height. In this case, if we use such a data set as it is, it may bias the system to tall persons  and there by the diagnosis of tall patients is likely to be more accurate  than patients with short height. This will make the decision making system "unfair" to people with short height.  We do not have the  vital information such as height (note that height is just an example, there could be many other factors) about the patients, yet we want to develop a scheme so that such bias can be avoided and at the same time we can maintain the desired level of variation in the instances used for designing the system.  Keeping this in mind, we propose an image selection algorithm named Variation Preserving Equally Distributed Image Selection (VarPEDIS).

We have used the DenseNet-121 \cite{FShi2020} pretrained with ImageNet  to generate  raw features from the CXR images - a vector of length 1024 for each image. DenseNet-121 was chosen because it works well on chest X-rays as mentioned by Rajpurkar et al. (2017).  Suppose we have $n_i$ images in the CXR data set for class $i$. Thus, from the DenseNet-121, for the $i^{th}$ class we shall get 
$n_i$ feature vectors $\textbf{X}_i=\{ \textbf{x}_1, \textbf{x}_2, ..., \textbf{x}_{n_i}\} \subset R^{1024}$. Let $\textbf{v}_i$ be the centroid of those $n_i$ embedded feature vectors in $\textbf{X}_i$. We compute the similarity of an image (say $j^{th}$ image of the $i^{th}$ class) with the centroid $\textbf{v}_i$ as  

\begin{equation}\label{similarity}
s_j=\frac{<\textbf{x}_j, \textbf{v}_i>}{||\textbf{x}_j|| ~ ||\textbf{v}_i||} \;\; \forall j=1, 2, \cdots, n_i.
\end{equation}

We note that one can use other measures of similarity also. If for an image, the similarity is less than a threshold, we discard that image for further processing. This process helps us to discard very poor quality (poor in terms of representing the associate class) images. 

Now we distribute the images into a set of equal-width buckets using the measure of similarity in Eqn. (\ref{similarity}).  If a bucket has less than 200 instances we adjust the bucket width to accommodate at least 200 instances. Finally, from each bucket we randomly select a fixed number of representative points. This process helps to maintain the diversity of the original image set into the selected set. It also reduces the training time. Moreover, for data sets like COVID-19, where we have very limited number of labeled instances, we can reduce the problem of learning with  highly imbalanced 
training data. The rational behind sampling evenly from equal-width buckets can be summarized as follows:

\begin{enumerate}[label=(\alph*)]
\item Every bucket may be viewed as representing a true repeatable situation.
\item  There are many methods to deal with the explicit imbalance between classes in terms number of instances in each class. But here we are trying to deal with 
some hidden imbalances within a class that can make a system unfair.  There could be many causes for making such imbalance; for example, it could be age distribution or the gender distribution in the  population for a particular class. The decision system should not be influenced by such hidden biases.
\item Classifier should learn all valid variations both within a class and between classes.
\end{enumerate}

Algorithm \ref{alg:VarPEDIS} describes the detailed steps involved in the image selection process.

\begin{figure*}
  \includegraphics[width=\textwidth,height=7cm]{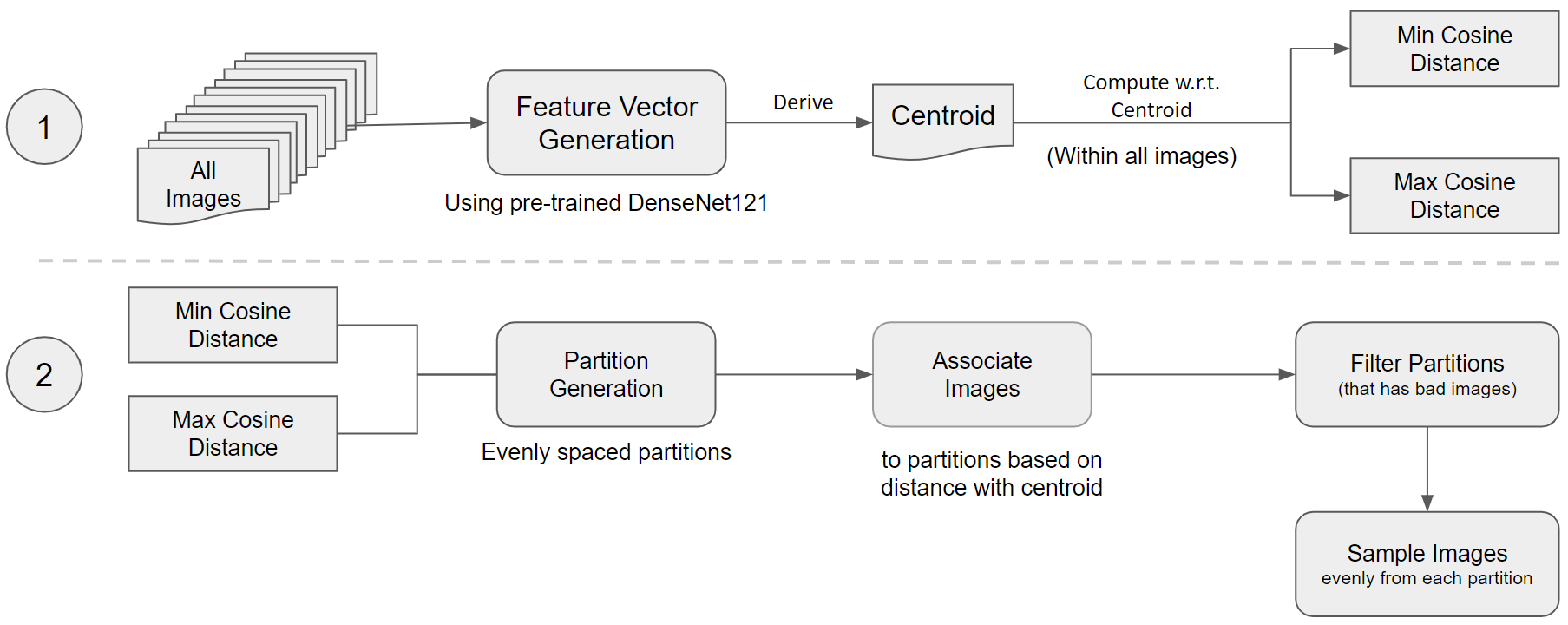}
  \caption{Maximum Variation Image Selection Workflow}
  \label{fig:network}
\end{figure*}

\begin{algorithm} 
\caption {VarPEDIS: Variation Preserving Equally Distributed Image Selection Algorithm}
\label{alg:VarPEDIS}
\begin{itemize}[leftmargin=1.2cm]
\item [Step 0:] Input $: \textbf{X}=\cup_{i=1}^c \textbf{X}_i; \textbf{X}_i \cap \textbf{X}_j= \Phi; i\ne j; |\textbf{X}_i| =n_i; \sum_{i=1}^c n_i=N$, $\textbf{X}_i$ is training samples from class $i$. \;
\item [Step 1:] If the number of images in a class, $n_i \le 500$, we select them all; Otherwise,
\begin{itemize}[leftmargin=0.5cm]
\item [Step 1.1] Feature embedding using the pretrained DenseNet-121
\item [Step 1.2] Compute the centroid of  the embedded representations of all instances in a class.
\item [Step 1.3] Compute Cosine similarity between the embedded representation of the   images and the centroid of  the corresponding class.
\item [Step 1.4] Select only images with similarity greater than a threshold, $\theta$.
\item [Step 1.5] Distribute the selected images into $K$ buckets of equal width. If a bucket does not have at least $N$ images, we adjust the bucket width to have $N$ images.
\item [Step 1.6] From each bracket randomly select $N1$ images leading to a sample of size only $K.N1$ for a class.
\end{itemize}
\item [Step 2] Use the Lung-Finder in \cite{LungsFinder} to locate the lung areas in each of the selected images.
\item [Step 3] Crop the smallest rectangle containing the lung segments and resize it into an image of size $224 \times 224$.
\end{itemize}
\end{algorithm}
Note that, in Algorithm \ref{alg:VarPEDIS}, we have chosen $k=5$, $N$=$N1=200$. The number of buckets and the number of instances to be selected from each bucket may need to be changed depending on the training data set. Fig. \ref{fig:network} gives  a pictorial representation of the workflow for the variance preserving image selection.

\begin{figure}[!htb]
	\begin{center}
		\scalebox{0.17}[0.17]{\includegraphics{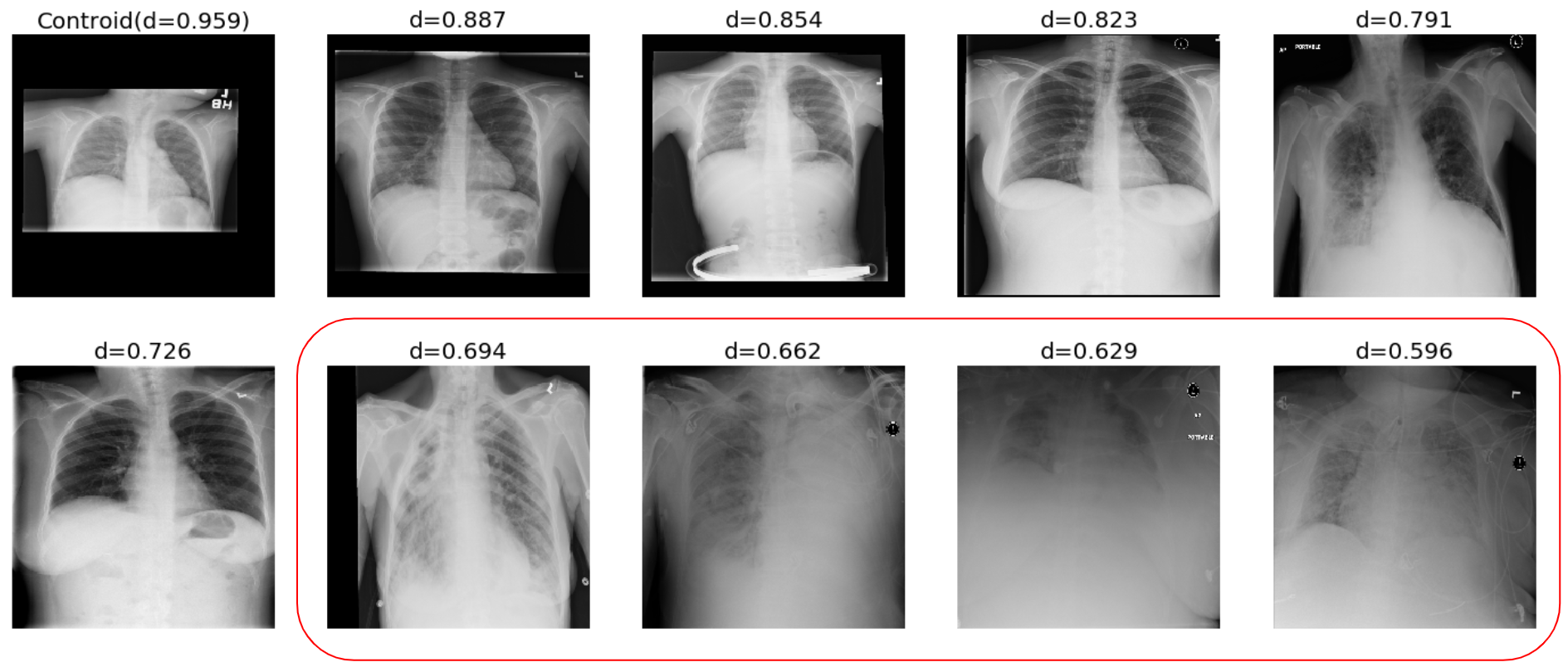}}
	\end{center}
	\caption{Infiltration CXRs, images in red block are ignored. d represents cosine similarity with centroid }
	\label{fig:SelectedRejected}
\end{figure}

In Fig. \ref{fig:SelectedRejected} we depict six selected images and four discarded images from the class Infiltration. The first image in the upper 
panel is the image closest to centroid and as expected its similarity value is very high. The discarded images are selected using a threshold $\theta=0.7$. Fig. \ref{fig:SelectedRejected} reveals that the discarded images are indeed poor quality images.

Now using the selected instances, we use a two-phase training as summarized in Algorithm \ref{alg:twophasetraining}.


\begin{algorithm}
\caption{Two-Phase Training}
\label{alg:twophasetraining}
\begin{itemize}[leftmargin=1.1cm]
\item[Step 1] Input: The training data selected by Algorithm \ref{alg:VarPEDIS}
\item [Step 2]  Phase One of  training
\begin{itemize}
	\item [Step 2.1] Train only the last layer of the Dense Net using Stochastic Gradient Descent (SGD) for a limited number of epochs using the entire training data.
\end{itemize}
\item [Step 3]  Phase Two of  training
\begin{itemize}
	\item [Step 3.1] Train the full network using the training data for all 15 classes using Adam optimizer.
\end{itemize}
\end{itemize}
\end{algorithm}

%
%
%
%
%

%
%
%
\section{The Network Model and its Training}\label{SecModel}

\begin{figure}[!b] 
	\begin{center}
		\scalebox{0.3}[0.3]{\includegraphics{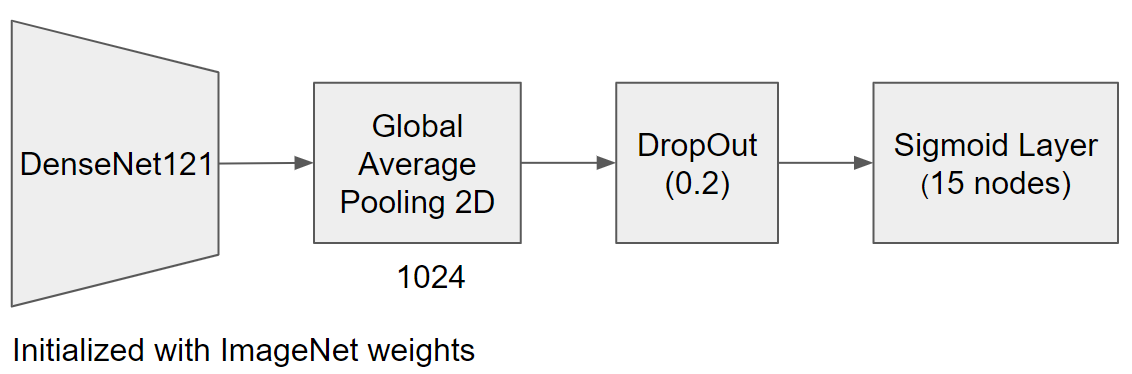}}
	\end{center}
	\caption{The Classification Model}	
	\label{fig:model}	
\end{figure}

As mentioned earlier, we have used DenseNet-121 since several other studies suggested its choice for CXR image classification.
We have also obtained, as we shall see soon,  very good results. Fig. \ref{fig:model} depicts  the overall network model.

DenseNet-121 was initialized with ImageNet weights. In the 1st phase of training, only the last layer was trained using Stochastic Gradient Descent (SGD) optimizer (LR - 0.001, Nesterov momentum - 0.9). A maximum of 10 epochs were allowed in this phase. In the final phase of training the entire network was trained using Adam optimizer( LR = 0.0001, beta1 = 0.9,  beta2 = 0.999). We use the early-stopping criterion monitoring the validation error. In particular, the network with the minimum validation error, when there was no improvement in the validation performance over 7 epochs, is chosen for testing.

As described in Algorithm \ref{alg:VarPEDIS}, the lung region was extracted from the CXR images by combining the lung segment rectangles identified by lung\_finder \cite{LungsFinder}. 
Images were resized to $256 \times 256$ initially and finally transformed to $224 \times  224$ because the default input size to DenseNet-121 is $224 \times  224$.

In order to realize a robust system, we have used the data augmentation features available with TensorFlow. In fact, we have used Width\_Shift, Height\_Shift, Rotation, Brightness change, and Horizontal\_flip. 

%

\section{Experimental Results}\label{SecExperimentalResults}
In our experiments, we measured the performance of our system at classifying 15 different lung diseases including COVID-19 and found very consistent 
performance in classifying COVID-19. 

In order to demonstrate how effective the proposed method is, we compare the performance of our method with four other methods 
as reported in Table \ref{table:resultstable} on page \pageref{table:resultstable}. Note that, in \cite{PranavRajpurkar2017} only 14 types of lung diseases have been classified but in our case we consider an additional class, COVID-19, i.e., 15 classes and that makes the problem more difficult but this scenario is much more natural from a practical application point of view. 
Since we have used the ChestX-ray14 dataset as well as the COVID-19 data set of Cohen \cite{JPCohen2020}, the experimental protocols 
may not be exactly the same with those of the compared methods. Yet, the comparison is very useful and clearly demonstrates the effectiveness of the proposed system. In our experiment, we have used 13,599  images for training, validation, and testing. The performance of the proposed model is evaluated with 5-fold cross-validation. We have repeated the experiment 10 times and the average results are reported.

Table \ref{table:resultstable} on page \pageref{table:resultstable} summarizes the comparison of our methods with the four other methods in terms of AUROC. Note that, the results in the first four columns are directly taken from \cite{PranavRajpurkar2017}. The last column depicts the average performance of the proposed methods over different classes including COVID-19. A careful inspection of the Table \ref{table:resultstable} reveals that of the 14 non-COVID-19 classes, our method produced the best results in six cases which are highlighted in bold. More noticeably, for the COVID-19 cases, the performance score is 0.956 which is higher than the performance score of any of the four methods for any of the 14 categories of diseases.

\begin{figure}[h]
	\begin{center}
		\scalebox{0.15}[0.15]{\includegraphics{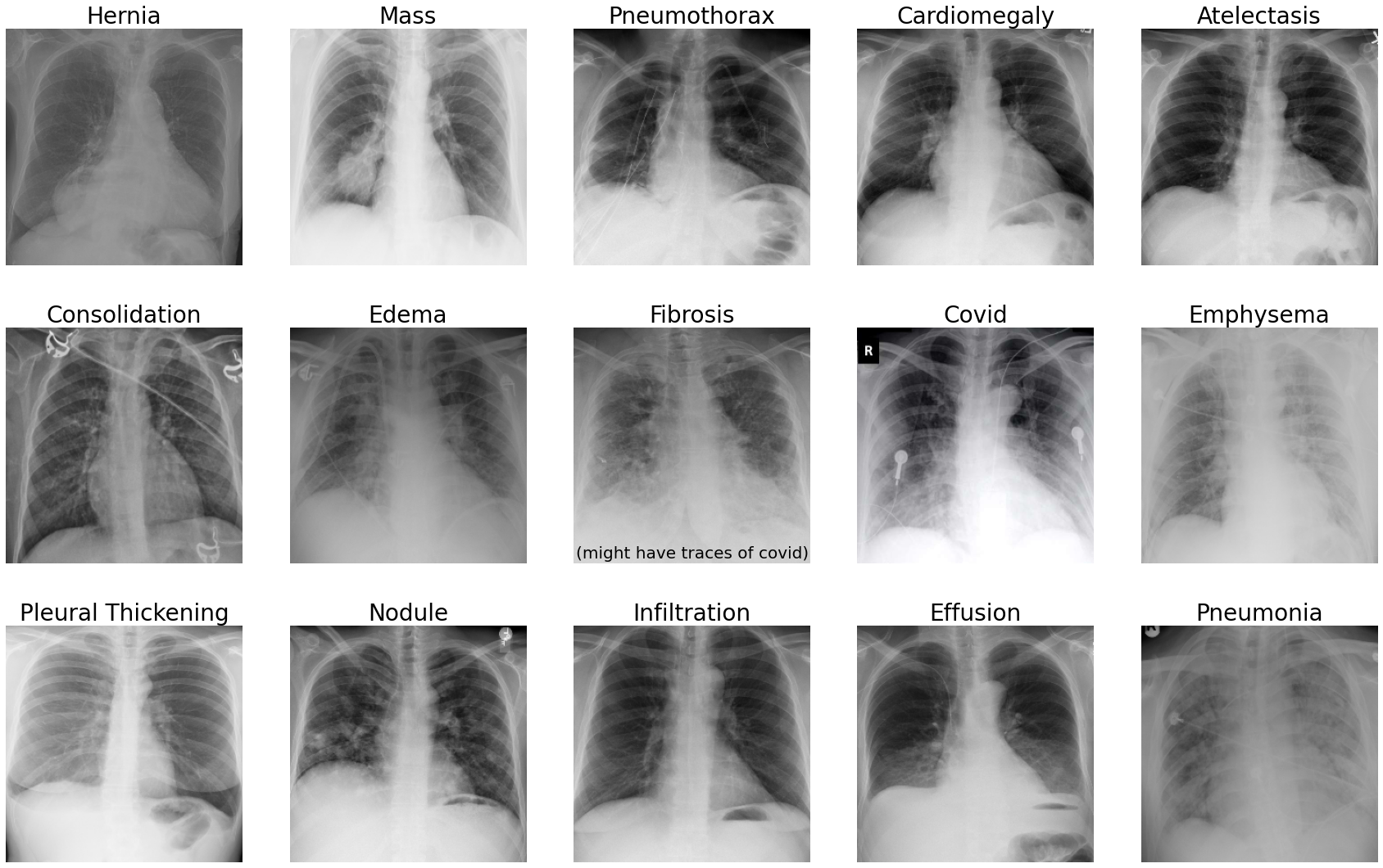}}
	\end{center}
	\caption{Examples of correct classification including COVID-19 }
	\label{fig:correctclassification}
\end{figure}

In Fig. \ref{fig:correctclassification}, as illustrations, we display one example each of the 15 classes which is correctly classified by the system.  

\begin{figure}[h]
	\begin{center}
		\scalebox{0.3}[0.3]{\includegraphics{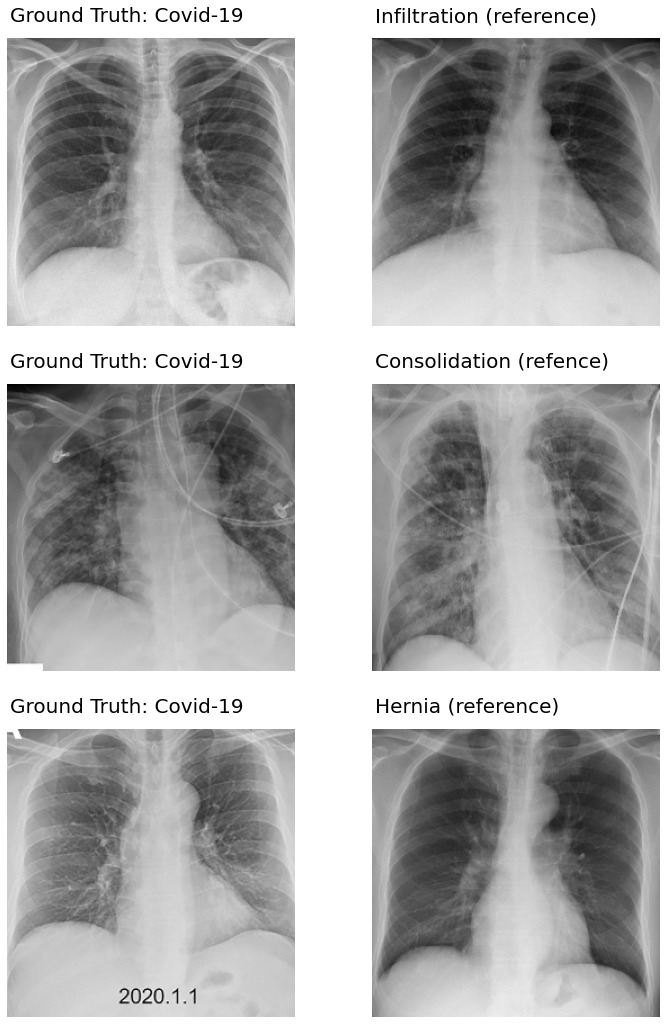}}
	\end{center}
	\caption{Some COVID-19 misclassifications }
	\label{fig:covid19misclassified}
\end{figure}

Next in Fig. \ref{fig:covid19misclassified} we display a  few COVID-19 cases which are incorrectly classified by the proposed method. The first column depicts the COVID-19 images which are misclassified.  The first image (in the top-left panel) is misclassified as Infiltration. 


\begin{table*}[t]
\begin{tabular}{ |m{2.5cm}|m{2.5cm}|m{2.5cm}|m{2.5cm}|m{2.5cm}|m{2.5cm}| }
\hline
\thead{Disease} & \thead{ChestX-ray14 \\ Wang et al. (2017) \\ \cite{XiaosongWang2017}\cite{JosephPaulCohen2020} \\ DenseNet-50} & \thead{CheXNet \\ Rajpurkar et al. (2017) \\ \cite{PranavRajpurkar2017}\cite{JosephPaulCohen2020} \\ DenseNet-121} & \thead{CheXNet-Py3 \\ Weng et al. (2017) \\ \cite{XinyuWeng2017}\cite{JosephPaulCohen2020} \\ DenseNet-121} & \thead {Chester \\ Cohen et al. (2020) \\ \cite{JosephPaulCohen2020} \\ DenseNet-121} & \thead {Proposed Method} \\
\hline
Atelectasis & 0.71 & 0.80 & 0.81 ± 0.01 & 0.84 ± 0.01 & 0.80 ± 0.02 \\
\hline
Cardiomegaly & 0.80 & 0.92 & 0.90 ± 0.01 & 0.92 ± 0.01 & \textbf{0.94 ± 0.01} \\
\hline
Effusion & 0.78 & 0.86 & 0.87 ± 0.01 & 0.88 ± 0.01 & 0.83 ± 0.01 \\
\hline
Infiltration & 0.60 & 0.73 & 0.70 ± 0.01 & 0.73 ± 0.01 & 0.72 ± 0.03 \\
\hline
Mass & 0.70 & 0.86 & 0.82 ± 0.01 & 0.87 ± 0.01 & 0.84 ± 0.01 \\
\hline
Nodule & 0.67 & 0.78 & 0.74 ± 0.01 & 0.79 ± 0.01 & \textbf{0.82 ± 0.01} \\
\hline
Pneumonia & 0.63 & 0.76 & 0.76 ± 0.02 & 0.72 ± 0.04 & \textbf{0.76 ± 0.02} \\
\hline
Pneumothorax & 0.80 & 0.88 & 0.83 ± 0.01 & 0.86 ± 0.01 & 0.86 ± 0.02 \\
\hline
Consolidation & 0.70 & 0.79 & 0.79 ± 0.01 & 0.81 ± 0.01 & 0.77 ± 0.02 \\
\hline
Edema & 0.83 & 0.88 & 0.86 ± 0.01 & 0.91 ± 0.01 & \textbf{0.92 ± 0.01} \\
\hline
Emphysema & 0.81 & 0.93 & 0.89 ± 0.01 & 0.93 ± 0.01 & 0.86 ± 0.01 \\
\hline
Fibrosis & 0.76 & 0.80 & 0.78 ± 0.01 & 0.78 ± 0.01 & \textbf{0.82 ± 0.01} \\
\hline
Pleural Thickening & 0.70 & 0.80 & 0.75 ± 0.01 & 0.81 ± 0.01 & 0.74 ± 0.02 \\
\hline
Hernia & 0.76 & 0.91 & 0.88 ± 0.03 & 0.83 ± 0.07 & \textbf{0.96 ± 0.03} \\
\hline
Covid & None & None & None & None & 0.96 ± 0.01 \\
\hline
 &  &  &  &  &  \\
\hline
Average & 0.73 & 0.84 & 0.81 & 0.83 & 0.84 \\
\hline
\end{tabular}
\caption{Comparison of the proposed method in classifying COVID-19 along with 14 other lung diseases with results from the literature where the compared method did not consider COVID-19.  In the last column, cells with bold face entries indicate better performance.}\label{table:resultstable}
\end{table*}

The image in the second panel of row one displays the training image  from the Infiltration  class that has the highest similarity with the COVID-19 image. Inspection of these two images reveals that both have similar opacification in the lung areas.

The images in the first column of second and third rows are the  other two COVID-19 instances which are misclassified as Consolidation and Hernia, respectively. The second image in row two is the training image from the consolidation training set having the highest similarity with the COVID-19 image.
Comparison of the first and second images in the second row reveals that both have consolidations on the lungs towards the center of the images, which probably is the cause of misclassification.

%
%

\section{Conclusions}\label{SecConclusions}
The COVID-19 pandemic has shaken the world. Many countries are still struggling with the second wave and a third wave is also expected. 
Time demands a low cost and quick detection method for COVID-19. Consequently, diagnosis of COVID-19 using Chest X-Ray images has become a hot topic of research. 
The majority of the research efforts are restricted to classification of COVID-19 as a problem involving 2-4 classes. More importantly, adequate and explicit emphasis has not been given to make the methods unbiased, where the bias could be caused by data imbalance between different classes or even due to hidden factors like  gender imbalance, age variation, among others within a class of the  training data. 
Two major achievements in this study are: First, we have considered 15 lung diseases along with COVID-19 which is a much more difficult and realistic problem to solve. 
Second, without knowing any patient specific information except the CXR images, we have developed a system that tries to minimize biases in decision making. This is achieved using a novel  instance selection scheme which selects instances for the training preserving the useful variation in the data set and discards instances that might cause problems during the training. The method also deals with hidden biases within a class.  We have used DenseNet-121 for feature extraction and for instance selection a class is represented by a prototype, which could be computed in different ways such as the data point with the highest density or the centroid, we have used the centroids in this study.  The instance selection/rejection is done using a similarity measure. To demonstrate the effectiveness of the proposed method we have compared its performance with some published results which deals with simpler problems.  Yet, the proposed method obtained comparable or  better results.
 We are hopeful that the proposed technique or variants of this technique might yield good results for other image classification tasks. 
We also hope traces of COVID-19 type lung conditions might be present in pre COVID-19 CXRs. Identifying those might help determine treatment 
that has potential to become effective. There are some potential ways to improve further. For example, we have represented a class using one centroid. However,  
use of multiple centroids may help. We have not experimented with different choices of number of buckets. This may also improve the performance. Other measures of similarity measures may also be explored. 



%
%
%

\end{document}